\documentstyle[12pt,epsf]{article}
\def\vk{{\bf k}_\perp{}}

\begin{document}
\begin{center}
{\bfseries  ON THE $\sigma _L/\sigma _T$ RATIO IN
POLARIZED VECTOR MESON PHOTOPRODUCTION}

\vskip 5mm

S.V.Goloskokov$^{\dag}$

\vskip 5mm

{\small {\it Bogoliubov Laboratory of Theoretical  Physics,
 Joint Institute for Nuclear Research,
Dubna 141980, Moscow region, Russia}
\\
$\dag$ {\it E-mail: goloskkv@thsun1.jinr.ru }}
\end{center}

\vskip 5mm

\begin{center}
\begin{minipage}{150mm}
\centerline{\bf Abstract} We study the spin-dependent
cross-sections of vector meson photoproduction for longitudinally
and transversely polarized photons within a QCD- model. The
dependence of  the $\sigma _T/\sigma _L$ ratio on the photon
virtuality and on the meson wave function is analysed.
\end{minipage}
\end{center}

\vskip 10mm

Investigation of diffractive vector meson photoproduction is a
problem of considerable interest now. The factorization of
diffractive vector meson production with longitudinally polarized
photons  into the hard part and parton distribution was proved in
\cite{coll}. Thus, on the one hand, such processes, can provide an
excellent tool to study the parton distribution in a hadron at
small $x$. On the other hand, they should give an important
information on the vector meson wave function. One of the
observables which was investigated  experimentally at HERA
\cite{rr} is the ratio of the cross sections with transversely
and longitudinally polarized photons $R=\sigma _L/\sigma _T$ for
nonzero photon virtuality.  To calculate $R$, the amplitudes with
transverse vector meson polarization within some model approach
should be analysed. The nonrelativistic wave function gives the
ratio $R \sim Q^2/M_V^2$ \cite{rys1}which is not supported by the
experiment. More realistic wave functions which include the
transverse quark motion can improve the situation \cite{rys2,cud}.

In this report, we focus our attention on  dependences of the $R$
ratio on the form of the meson wave function. We consider the
following matrix structure  of the vector meson wave function
\begin{equation}
\label{wf1}
 \hat \psi_V=g \frac{(\hat k_1+ m_q) \hat E_V (\hat k_2+ m_q)}{M_V z
(1-z)}\phi_V(z,k_\perp^2).
\end{equation}
Here $k_1$ and $k_2$ are the quark momenta, $E_V$ is the vector
meson polarization and $\phi_V(z,k_\perp)$ is the distribution
amplitude. The momentum $k_1$ carries the fraction $z$ of the
vector meson momentum and can be written in the form
\begin{equation}\label{k1}
k_1=-z V+ \Delta -K,
\end{equation}
where $K=[0,0,\vk]$ is the quark transverse momentum and the
vector $\Delta$ puts the quark momentum on the mass shell
$k_1^2=m_q^2$ in calculating the imaginary part of the amplitude.
The wave function (\ref{wf1}) can be decomposed into the following
structures
\begin{eqnarray}
\label{wf_m} \psi_V&=&g  \left[ \hat E_V \left( M_V
+\frac{m_q^2}{M_V z (1-z)}+ \frac{m_q}{M_V z (1-z)}\hat V
-\frac{K\cdot K}{M_V z (1-z)}
-\frac{2 (1- z) \Delta \cdot V}{M_V z (1-z)} \right) \right .\nonumber\\
&+&\frac{\hat V \hat E_V \hat K}{M_V z (1-z)} - \frac{2 E_V \cdot
K}{M_V z (1-z)} \left( (1-z) \hat V - \hat K +m_q
\right) \nonumber\\
&+&  \left. \frac{2 \Delta \cdot E_V}{M_V z (1-z)} \left( (1-z)
\hat V - \hat K + m_q \right)  \right] \phi_V(z,k_\perp^2).
\end{eqnarray}
We find that for small $m_q^2/M_V^2$ and $K^2/M_V^2$ the function
(\ref{wf_m}) reduces to the standard form of the vector meson wave
function $g \hat E_V ( M_V+\hat V)$. The same is true in the
nonrelativistic limit.

The leading term of the amplitude of diffractive vector meson
production is mainly imaginary.  The imaginary part of the
amplitude can be written as an integral over $z$ and 2-dimensional
integrals over $k_\perp$ and $l_\perp$. The integral over
$l_\perp$ can be represented in terms of the skewed gluon
distribution ${\cal F}^g$ and has the form \cite{goljp}
\begin{eqnarray}\label{btn}
\sum T_i &\propto&\frac{1}{\left( k_\perp^2+|t|+\bar Q^2 \right)}
\int \frac{d^2l_\perp (l_\perp^2+\vec l_\perp \vec r_\perp-\vec
l_\perp \vec k_\perp) G(l_\perp^2,x,...)}
{(l_\perp^2+\lambda^2)((\vec
l_\perp+\vec r_\perp)^2+\lambda^2)[(\vec l_\perp-\vec k_\perp)^2+|t|+\bar Q^2]} \nonumber\\
&\sim& \frac{1}{\left( k_\perp^2+|t|+\bar Q^2
\right)^2}\int^{l_\perp^2<k_\perp^2+\bar Q^2+|t|}_0
\frac{d^2l_\perp (l_\perp^2+\vec l_\perp \vec r_\perp) }
{(l_\perp^2+\lambda^2)((\vec l_\perp+\vec r_\perp)^2+\lambda^2)}
G(l_\perp^2,x,...) \nonumber\\
&=&\frac{1}{\left( k_\perp^2+|t|+\bar Q^2 \right)^2} {\cal
F}^g_{x}(x,t,k_\perp^2+\bar Q^2+|t|),
\end{eqnarray}
where $\bar Q^2=m_q^2+z (1-z)Q^2$, $G$ is the nonintegrated gluon
distribution, and $r_\perp$ is the transverse part of the momentum
carried by the two-gluon system. The distribution ${\cal
F}^g_{0}(x,0,q_0^2)$ is normalized to $(x g(x, q_0^2))$.

We use the helicity conservation hypothesis in diffractive vector
meson production. In this case one should calculate only two
amplitudes with longitudinal $A_{L}= A_{\gamma_L \to V_L}$ and
transverse $A_{T}= A_{\gamma_T \to V_T}$ photon polarization.
Other amplitudes which do not conserve helicity vanish as $|t|
\to 0$. The following approximation for the skewed gluon
distribution
\begin{equation}\label{spd}
{\cal F}^g_{x}(x,t,q_0^2)\simeq F(t)(x g(x, q_0^2)),
\end{equation}
is used for simplicity, where $F(t)$ is a hadron form factor. The
longitudinal amplitude is found to be as follows
\begin{eqnarray}
\label{Al} A_L&=&4\,N\,\int dz \int d k_\perp^2 (x g(x,
k_\perp^2+\bar Q^2+|t|)) \phi_V(z, k_\perp^2)
\nonumber\\
&\cdot&{\frac {\sqrt {{Q^2}}\, \left ({ k_\perp^2}+{{ m_q}}^{2}+
{{ M_V}}^{2}{ z (1-z) }\right )\left ({ k_\perp^2}-{ \bar
Q^2}\right )} {{ M_V^2}\,\left ({ k_\perp^2}+{\bar Q^2}\right
)\left( k_\perp^2+|t|+\bar Q^2 \right)^2}}.
\end{eqnarray}
For the transverse one we have
\begin{eqnarray}
\label{At} A_T&=&2\,N\,(E_V^T \cdot E_\gamma^T)\int dz \int d
k_\perp^2 (x g(x, k_\perp^2+\bar Q^2+|t|)) \phi_V(z, k_\perp^2)
\nonumber\\
&\cdot&{\frac {{ k_\perp^2}\,{{ m_q}}^{2}-2\,{{ k_\perp^2}}^{2}{ z (1-z) }+2\,{
 k_\perp^2}\,{ z (1-z) }\,{ \bar Q^2}-{ \bar Q^2}\,{{ m_q}}^{2}-2\,{ k_\perp^2}\,
{ \bar Q^2}}{M_V  z (1-z) \left ({ k_\perp^2}+{ \bar Q^2}\right
)\left( k_\perp^2+|t|+\bar Q^2 \right)^2}}.
\end{eqnarray}
Here N is some normalization factor which is the same for
longitudinal and transverse amplitude. The gluon distribution in
the case of vector meson production depends on $x$ which is fixed
by $x=x_p=(Q^2+|t|+M_V^2)/W^2$ . In (\ref{Al},\ref{At}) we leave
a small $t$ variable in the denominator and in the scale of the
gluon structure function. Note that at HERA energies, we can
consider only the gluon contribution to the amplitudes
(\ref{Al},\ref{At}).

The cross sections with longitudinal and transverse photon
polarization are $\sigma_{L(T)} \propto A_{L(T)}^2$.  In our
analyses we use the same wave function $\phi_V$ for longitudinal
and transverse vector meson polarization which  vanishes as $z
(1-z)$ for $z \to 0$ or $z \to 1$. In this case it can be found
that for $t=0$ we have divergence in $\sigma_{T}$. This problem
can be solved in two separate ways. The first one was proposed in
\cite{rys2} for $|t|=0$ by considering the scale dependence in
the gluon distribution $x g(x,k_\perp^2+\bar Q^2) \propto
(k_\perp^2+\bar Q^2)^\lambda$ with $\lambda \sim 0.3$ for small
$x$. The additional factor $(k_\perp^2+\bar Q^2)^\lambda$ reduces
the divergence of the integral, and one can found that \cite{rys2}
\begin{equation}\label{sig}
\sigma_{L} \propto \frac{1}{(1+\lambda)^2}\;\;\;\;\;
\mbox{and}\;\;\;\;\; \sigma_{T} \propto \frac{1}{\lambda^2}.
\end{equation}

\phantom{.} \vspace{.5cm} \epsfxsize=9.6cm
\centerline{\epsfbox{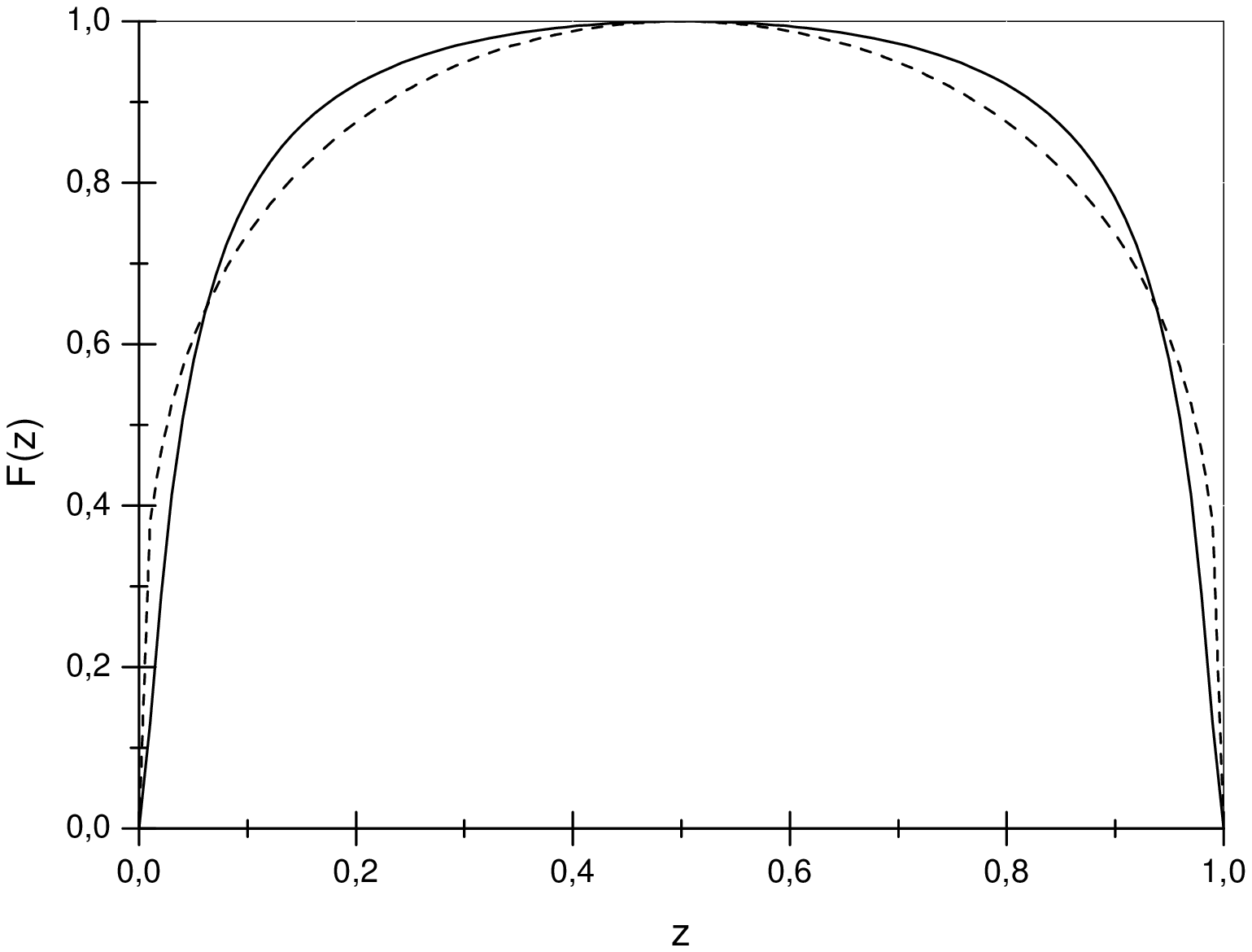}}
\begin{center}
Fig.1 Regularization functions at $Q^2=5 \mbox{GeV}^2$, normalized
to unity at $z=0.5$: $F^H$ -dashed line; $F^S$ -solid curve.
\end{center}

We call this method the "hard regularization". Some problem
appears here with the factorization of the amplitude to the hard
and soft part - gluon distribution becomes dependent on the
momentum $k_\perp^2$ of the hard contribution.

The other possibility is based on the fact that, in the fixed
target experiments it is usually impossible to determine $t$, and
the cross section integrated over momentum transfer
\begin{equation}\label{ds}
\sigma =\int dt \frac{d \sigma}{dt}
\end{equation}
is measured. The average momentum transfer can be estimated from
(\ref{ds}) to be about $|t| \sim 0.1 \mbox{GeV}^2$ for
(\ref{Al},\ref{At}). Thus, one can use the  "soft scale
regularization" which includes the nonzero momentum transfer $t$
in the denominator of (\ref{Al},\ref{At}) and in the gluon
distribution. In this case the integrals become convergent too.
As a result, we can write two different regularization functions
for integrals (\ref{Al},\ref{At}) at small $k_\perp^2$ and $m_q=0$
\begin{eqnarray}\label{rf}
F^H &=& \left( z(1-z) Q^2 \right)^\lambda \;\;\;\; \mbox{with}\; \lambda=0.3 \nonumber\\
F^S &=& \left( \frac{z(1-z) Q^2}{z(1-z) Q^2 +| \bar t|} \right)^2
\;\;\;\; \mbox{with} \; |\bar t| = 0.1 \mbox{GeV}^2.
\end{eqnarray}
The regularization functions normalized to unity are shown in
Fig.1. We see that $F^H$ practically coincides with $F^S$ at
$Q^2=5 \mbox{GeV}^2$. For larger $Q^2$ the difference between
these functions will be a little bit larger.

\vspace{1.1cm}
\epsfxsize=10cm \centerline{\epsfbox{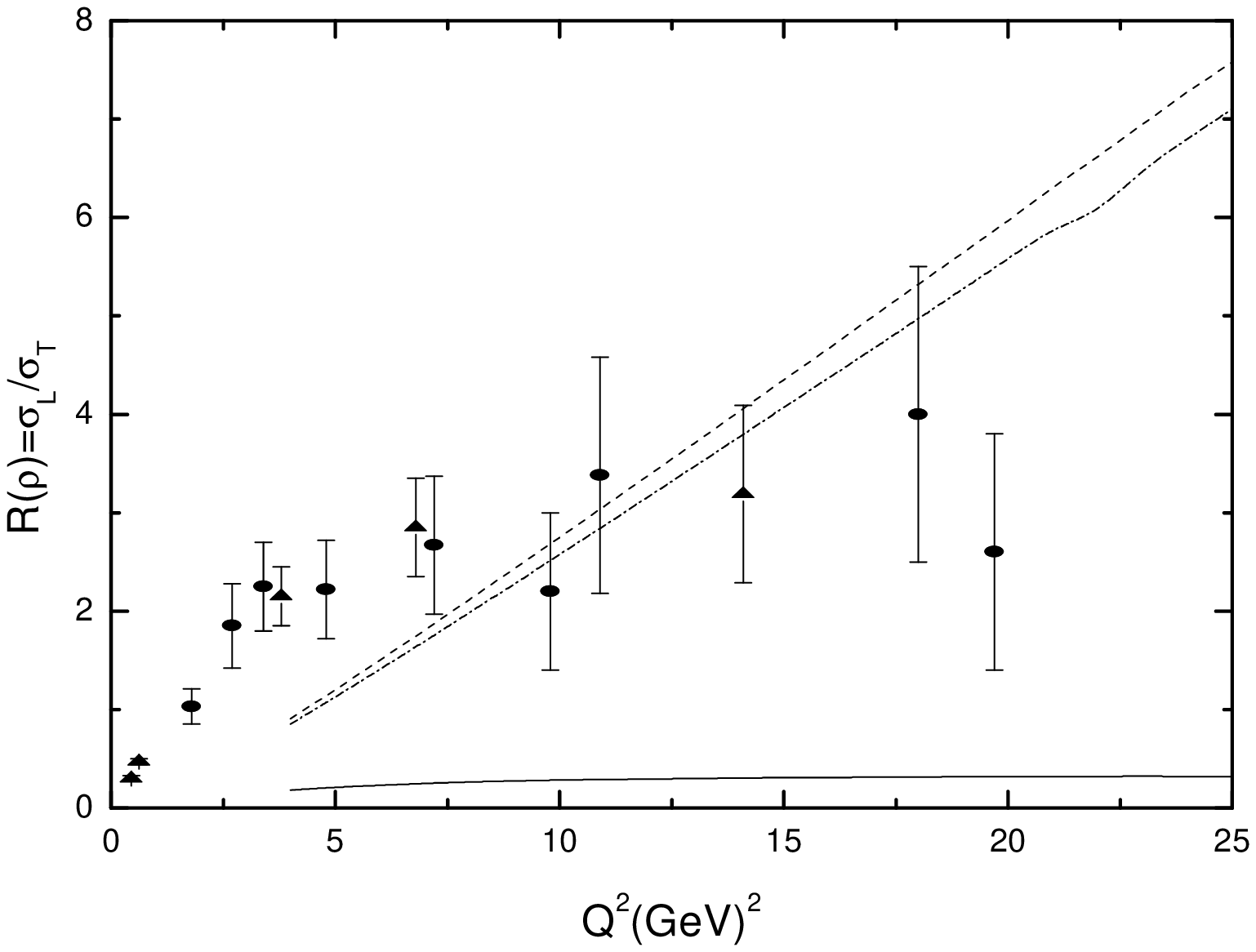}}
\begin{center}
Fig.2 $Q^2$ dependence of the ratio $R$ for $\rho$ production.
Solid curves - for $\phi_V^1$, dashed lines - for $\phi_V^2$,
dash-dotted lines - for $\phi_V^3$.
\end{center}

In what follows, we use the "soft regularization". The
corresponding cross sections $\sigma^{L,T} \propto (A_{L,T})^2$
were calculated for different forms of the wave function
$\phi_V(z, k_\perp^2)$  with the exponential $k_\perp^2$
dependence there \cite{kr}
\begin{eqnarray}
\label{wf}
\phi_V^1&=& N_1 z (1-z ) \exp{(-k_\perp^2 b g(z))} ,\; \;b=4.0[GeV^{-2}];\\
\phi_V^2&=&N_2 z (1-z ) g(z) \exp{(-k_\perp^2 b g(z))} ,\nonumber\\
&&g(z)=1/(z (1-z )),\;b=.88 [GeV^{-2}].
\end{eqnarray}
Here $N_i$ is a wave - function normalization factor which is
cancelled in the ratio of cross sections.

In addition, for $\rho$ production, the wave function that has a
two- maximum form in the $z$-distribution was tested
\begin{eqnarray}\label{wfm}
  \phi_\rho^3&=&N_3 z (1-z ) V(z) g(z) \exp{(-k_\perp^2 b g(z))},\nonumber\\
&&V(z)=1+.077 C_2^{3/2}(1-2 z)-.077 C_4^{3/2}(1-2 z),\nonumber\\
&&g(z)=1/(z (1-z )),\;b=.88 [GeV^{-2}].
\end{eqnarray}
This form of $V(z)$  was found from the QCD sum rules for the
$\rho$ meson in \cite{mikh}.

Calculations for $R=\sigma^{L}/\sigma^{T}$ were made for $\rho,
\phi\; \mbox{and}\; J/\Psi$ meson production for different $Q^2$
by using the same parameters $b$ in (\ref{wf}).

\vspace{1.cm}
\epsfxsize=10cm \centerline{\epsfbox{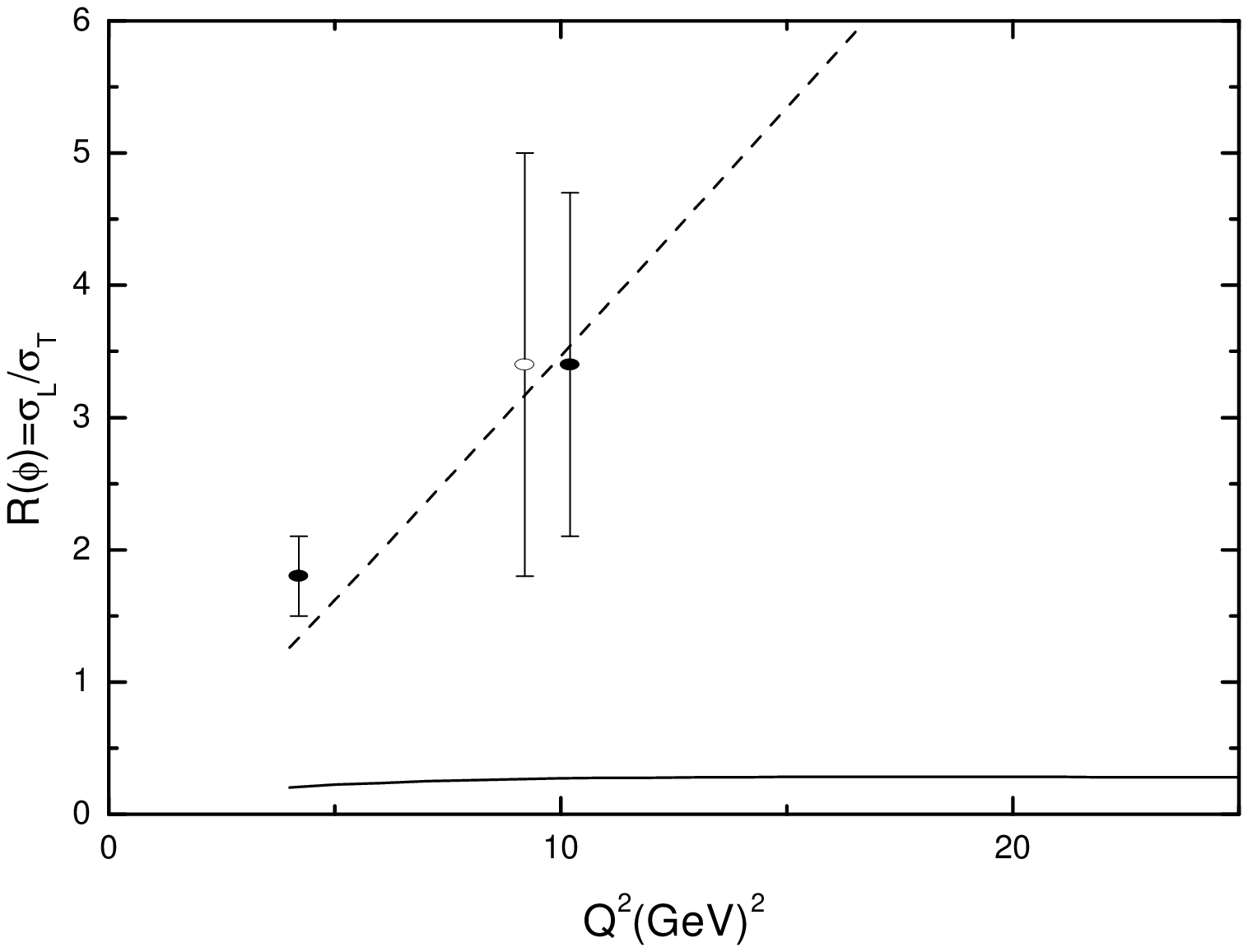}}
  \vspace*{.2cm}
\begin{center}
Fig.3 $Q^2$ dependence of the ratio $R$ for $\phi$ production.
Solid curves - for $\phi_V^1$, dashed lines - for $\phi_V^2$.
\end{center}

\vspace{.6cm}
\epsfxsize=10cm \centerline{\epsfbox{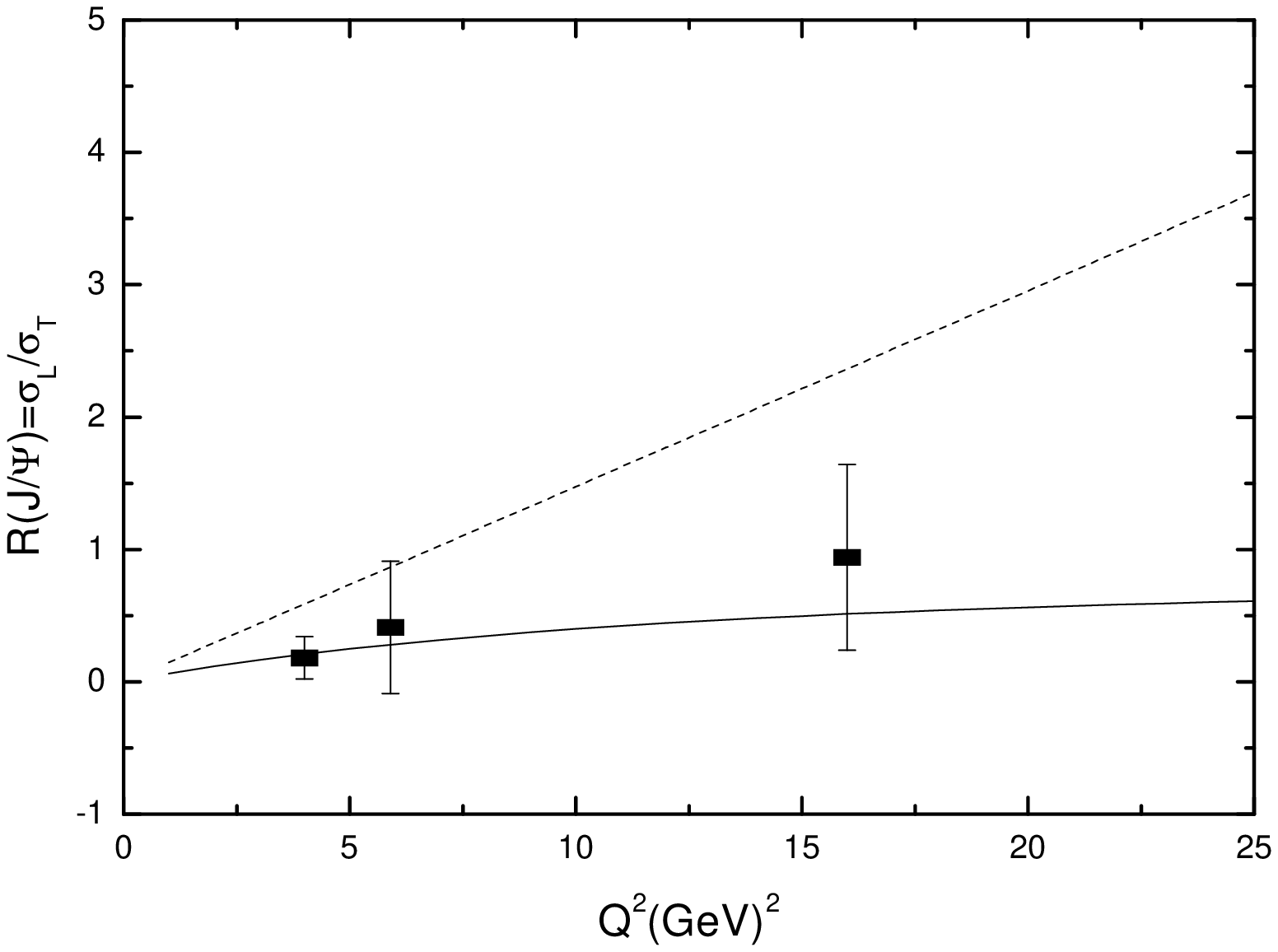}}
  \vspace*{.2cm}
\begin{center}
Fig.4 $Q^2$ dependence of the ratio $R$ for $J/\Psi$ production.
Solid curves - for $\phi_V^1$, dashed lines - for $\phi_V^2$.
\end{center}
Results for $\rho$ are shown in Fig. 2 together with known
experimental data. We find that the wave function  $\phi_V^1$
gives the $R$ ratio which  is essentially smaller than
experimental data. Results for the wave function $\phi_V^2$ are
compatible with data. At the same time, $R$ for the  wave function
$\phi_V^3$ is similar to that found for $\phi_V^2$. So, different
forms of z-dependences of the wave function give similar results
for the $R$ -ratio. However, this ratio is sensitive to the form
of the $k_\perp^2$ distribution of the wave function.

Results for $\phi$ meson production are shown in Fig. 3. They are
compatible with the data for $\phi_V^2$  as in the case of $\rho$
production. For $J/\Psi$ production, the wave functions $\phi_V^1$
and $\phi_V^2$ are not far from available experimental data
(Fig.4). This is caused by the nonzero quark mass $m_Q \sim
M_{J/\Psi}/2$ which regularizes integral for the $A_T$ amplitude.
Similar calculations were performed for the "hard regularization".
It was found that within 5-10 \% there is no difference between
the "hard" and "soft" regularization up to $Q^2\simeq 50-100
\mbox{GeV}^2$. Thus, we cannot determine now what physical
mechanism is more relevant to experiment.

To conclude, we would like to note that the effects of the gluon
distribution should be cancelled essentially in the ratio of cross
sections with different photon polarizations and the $R$ ratio is
expected to be dependent mainly on the vector meson wave function
structure. We find that the $R$ ratio is weakly dependent on the
$z$-distribution of the wave function. Really, results for the
asymptotic form $ \propto z(1-z)$  practically coincide (Fig.1)
with the wave function  found from the QCD sum rules $ \propto
z(1-z)V(z)$ which has two maxima in the $z$- distribution
($\phi_V^3$). At the same time, the $R$ ratio is strongly
sensitive to the transverse component of the wave function.
Really, for light quark production, the simple exponential form
$\exp{(-b k_\perp^2)}$ in the wave function gives a result which
is much smaller than experimental data. For the $z$- dependent
transverse component $\exp{(-b k_\perp^2/(z (1-z)))}$ the $R$
ratio is consistent with experiment. For heavy quark production,
the results of calculation for both wave functions (\ref{wf}) are
not far from experiment (Fig.4).

Note that, generally, the form of the vector meson wave function
for longitudinal and transverse polarization might be different
\cite{trwf}. This requires an additional investigation of the $R$
ratio in that case.

This work was supported in part by the Russian Foundation of Basic
Research, Grant 00-02-16696.

\end{document}